\documentclass[english,aps,prl,reprint,superscriptaddress]{revtex4-2}
\usepackage[T1]{fontenc}
\usepackage[utf8]{inputenc}
\usepackage{units}
\usepackage{graphicx}
\usepackage{hyperref}
\usepackage{babel}
\usepackage{amsmath}

\usepackage{color}
\usepackage[dvipsnames]{xcolor}
\usepackage[normalem]{ulem}

\begin{document}

\title{Experimentally measuring rolling and sliding in three-dimensional dense granular packings}

\author{Zackery~A.~Benson}
\affiliation{Institute for Physical Science and Technology, University of Maryland, College Park, MD 20742, USA}
\affiliation{Department of Physics, University of Maryland, College Park, MD 20742, USA}

\author{Anton~Peshkov}
\affiliation{Department of Physics, University of Rochester, Rochester, New York, USA}

\author{Nicole~Yunger~Halpern}
\email{nicoleyh@umd.edu}
\affiliation{Institute for Physical Science and Technology, University of Maryland, College Park, MD 20742, USA}
\affiliation{ITAMP, Harvard-Smithsonian Center for Astrophysics, Cambridge, MA 02138, USA}
\affiliation{Department of Physics, Harvard University, Cambridge, MA 02138, USA}
\affiliation{Research Laboratory of Electronics, Massachusetts Institute of Technology, Cambridge, Massachusetts 02139, USA}
\affiliation{Center for Theoretical Physics, Massachusetts Institute of Technology, Cambridge, Massachusetts 02139, USA}

\author{Derek~C.~Richardson}
\affiliation{Department of Astronomy, University of Maryland, College Park, MD, 20742, USA}

\author{Wolfgang~Losert}
\affiliation{Institute for Physical Science and Technology, University of Maryland, College Park, MD 20742, USA}
\affiliation{Department of Physics, University of Maryland, College Park, MD 20742, USA}

\begin{abstract}
We experimentally measure a three-dimensional (3D) granular system’s reversibility under cyclic compression.  
We image the grains using a refractive-index-matched fluid, then analyze the images using the artificial intelligence of variational autoencoders. These techniques allow us to track all the grains’ translations and 3D rotations with accuracy sufficient to infer sliding and rolling displacements. Our observations reveal unique roles played by 3D rotational motions in granular flows. We find that rotations and contact-point motion dominate the dynamics in the bulk, far from the perturbation's source. Furthermore, we determine that 3D rotations are irreversible under cyclic compression. Consequently, contact-point sliding, which is dissipative, accumulates throughout the cycle. Using numerical simulations whose accuracy our experiment supports, we discover that much of the dissipation occurs in the bulk, where grains rotate more than they translate. Our observations suggest that the analysis of 3D rotations is needed for understanding granular materials' unique and powerful ability to absorb and dissipate energy.
\end{abstract}
\maketitle

Research into granular rearrangements impacts diverse fields---astronomy, mechanical and civil engineering, geology, and industry~\cite{Richardson:2005,Friedmann:2006,Shin:2002}---in which granular packings undergo cyclic forcing. Even small cyclic deformations can significantly change a granular material’s mechanical properties~\cite{Espindola:2012}. 
For example, shearing can increase a material’s yield stress~\cite{Losert:2000}. Furthermore, the dependence of a material's properties on the external forcing's history could influence irreversible events such as rock avalanches~\cite{Friedmann:2006}.

Increasing attention has been dedicated to quantifying local rearrangements of grains in response to cyclic forcing. Such rearrangements explain the physics of aging, memory formation~\cite{Paulsen:2014,Fiocco:2018,Benson:2021}, energy dissipation~\cite{Zhai:2019,Bandi:2018,Dietrich:1998}, reversibility~\cite{Peshkov:2019}, and other changes in bulk rheological properties. For instance, one can observe a material's memory of a forcing amplitude by cyclically perturbing the material and measuring the resulting changes in the particles' positions~\cite{Paulsen:2014,Benson:2021, RMP:2019}.

Most previous studies of granular materials focus exclusively on grains' positions. However, grains interact via frictional contacts that generate torques, which cause rotations that form half of the system’s degrees of freedom (DOFs) (three DOFs are translational, and three are rotational). Grains in contact with each other can share another degree of freedom, which consists of \emph{contact-point motions}: rolling, sliding, and twisting at contact points~\cite{Bagi:2004}. In models that include friction, rolling is gear-like motion; the two grains rotate in opposite directions. When the net torque on a grain exceeds the friction's threshold, sliding occurs. Sliding is dissipative, generating heat dependent on the contact force between the grains~\footnote{See the Supplemental Materials for the equations for calculating sliding and rolling displacements, which include~\cite{Bagi:2004, Kabsch:1976}}. Additionally, dense configurations of spheres can form collective rotations that involve no sliding~\cite{Stager:2016}. Such configurations can be engineered into complex ball-bearing states for robotics~\cite{Pires:2020}.

Rotations play an important role in the mechanics of granular flow~\cite{Jaeger:1996}. For instance, grains in contact with each other can rotate in ways that soften the bulk material~\cite{Iwashita:1998}. Rotations also play important roles in continuum granular-flow models, in which rotations are generated through velocity gradients~\cite{Muhlhuas:1987,Benson:2021}. Additionally, sliding causes frictional dissipation that dominates energy loss in dense configurations that support few collisions~\cite{Jantzi:2020,Bandi:2018,Zhai:2019}. Therefore, to understand changes in a granular material’s mechanical properties, one must track not only grains' positions, but also rotations and contact-point motions.

Advances in three-dimensional (3D) imaging---namely, refractive-index-matched scanning~\cite{Dijksman:2012} and X-ray tomography~\cite{Zhai:2019}---enable accurate measurements of particles’ positions and rotations. In previous experiments, we quantified rotations by tracking the orientation of one axis in a spherical grain~\cite{Harrington:2014,Peshkov:2019}. These studies revealed that one-dimensional rotation is irreversible under cyclic compression. However, one axis allows us to measure only two of the three rotational degrees of freedom, prohibiting the tracking of 3D rotations and contact-point motions.

In this Letter, we take our previous experiments to a new level, quantifying rotational dynamics of individual grains in a cyclically compressed granular material. Our experimental setup tracks all three axes of each grain’s rotations, and our artificial-intelligence-enhanced image analysis greatly increases the particle-tracking's accuracy. These improvements allow us to compute individual grains' slipping and rolling displacements.  We thereby study the microscopic dynamics that result from cyclic forcing. We find that rotations and contact-point motions dominate the bulk dynamics, penetrating further into the material than translations do. Unlike particle translations, which are mostly reversible, 3D rotations turn out to be irreversible. Similarly, sliding, which is dissipative, accumulates throughout the compression cycle. We study the frictional dissipation using numerical simulations whose accuracy our experiment supports. We uncover significant dissipation in the bulk, where grains are rotating and scarcely translating. Thus, measuring 3D rotations is critical for understanding emergent phenomena, such as irreversibility and dissipation, in granular matter.

\begin{figure}[t]
\includegraphics[width=0.9\columnwidth]{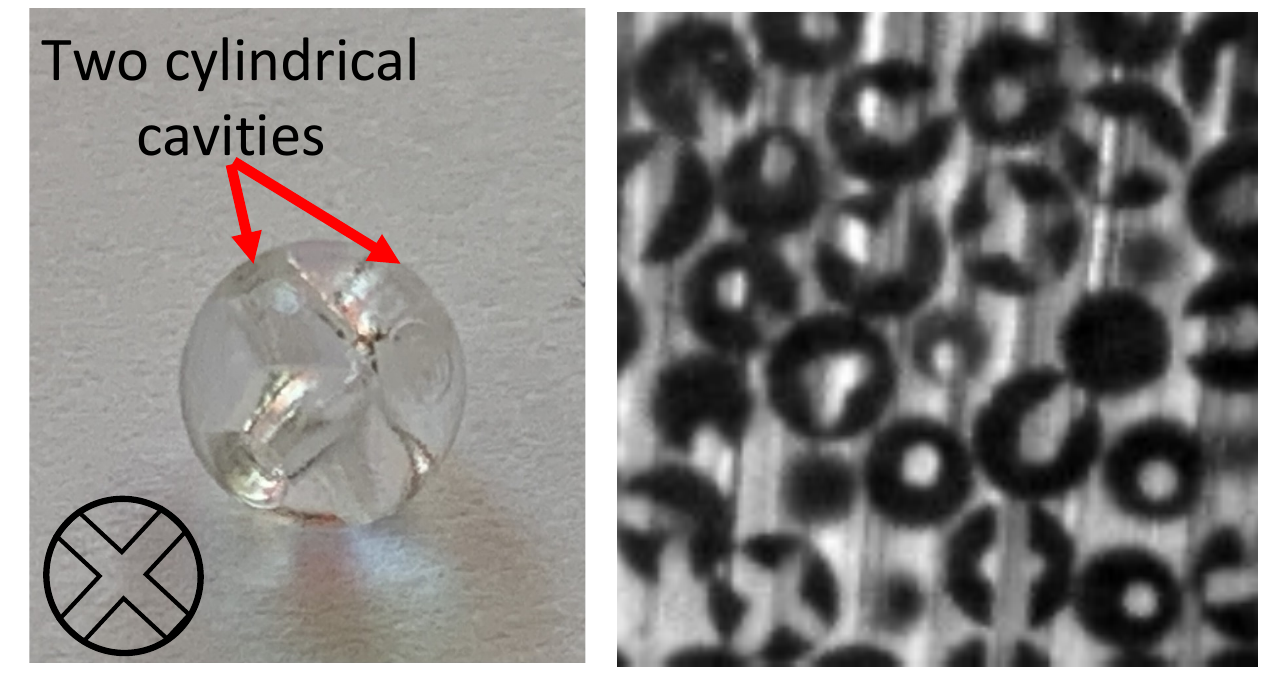}
\caption{\label{fig:one} Left: Photograph of an acrylic bead with two perpendicular cavities. Inset: Two-dimensional sketch of a bead, demonstrating how the two perpendicular cavities reveal the bead's orientation. Right: Sample $z$-slice of a few grains. Different spheres appear with different orientations. }
\end{figure}

\begin{figure}[b]
\includegraphics[width=1.0\columnwidth]{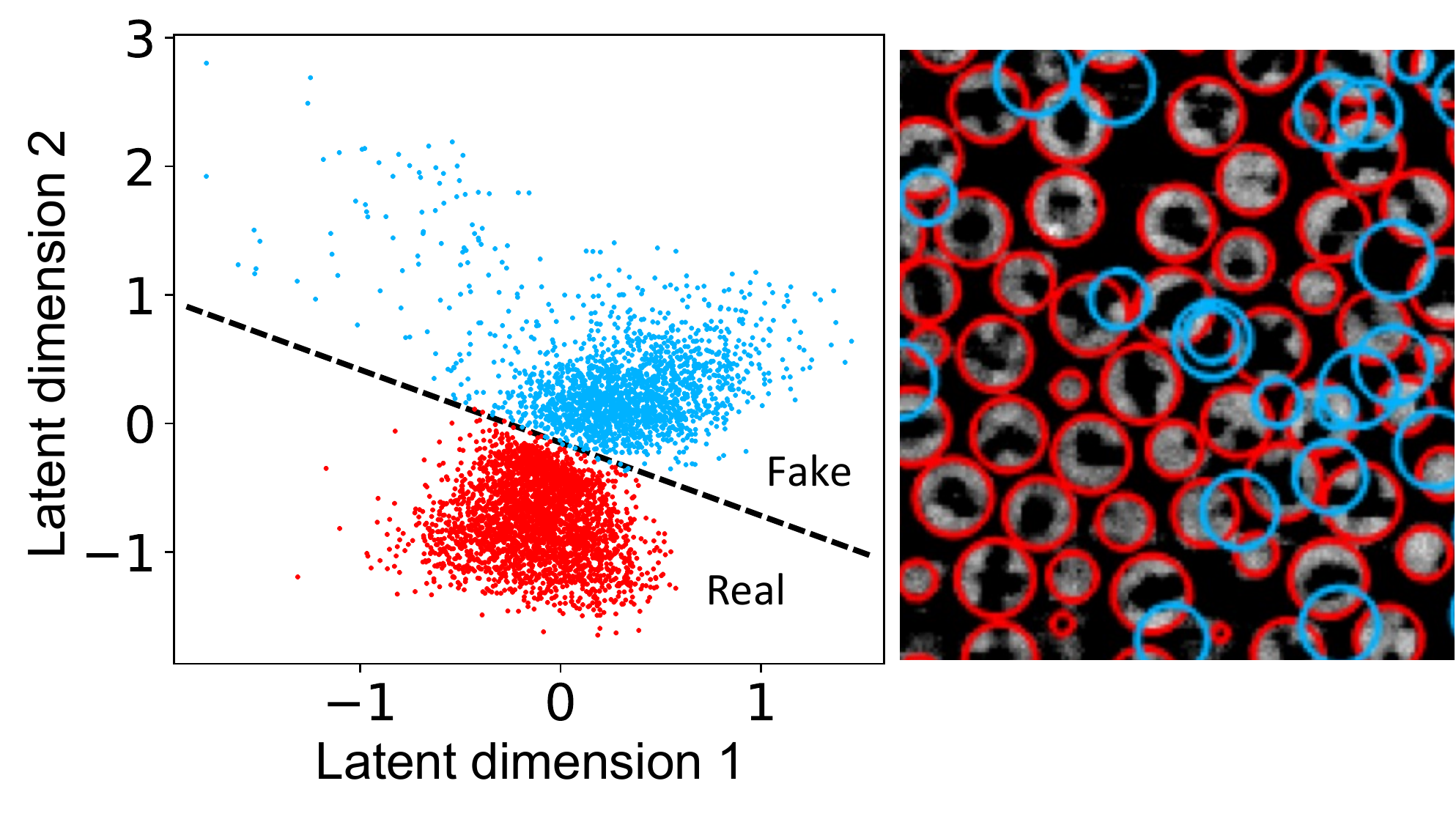}
\caption{\label{fig:two} Left: Latent-space points, produced by an artificial intelligence (a variational autoencoder), for all the grains in the sample. The colors demarcate real (red) from fake (blue) candidate grains. Right: Spatial positions of the candidate grains mapped to the latent-space dots.}
\end{figure}

\begin{figure*}[t]
\centering
\includegraphics[width=1.8\columnwidth]{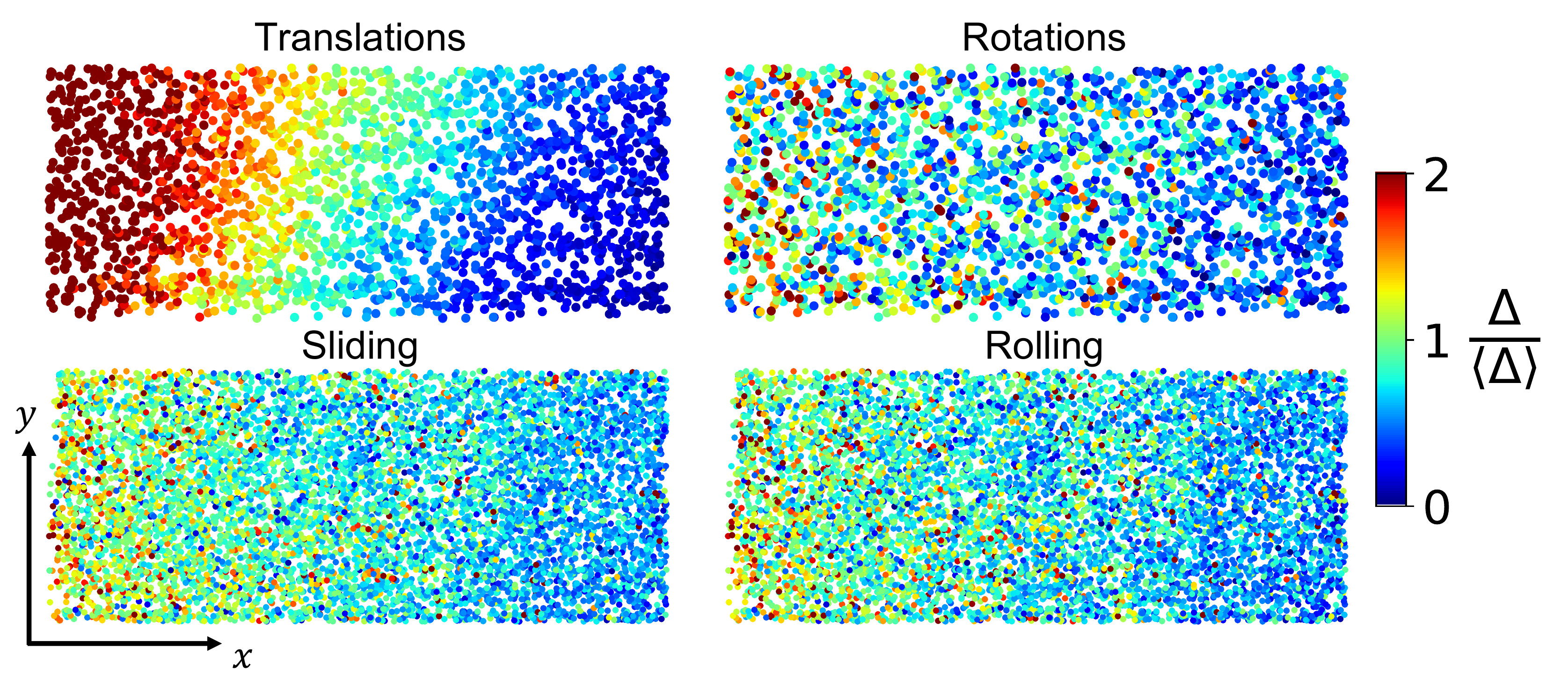}
\caption{\label{fig:three} Top row: Spatial distributions of grain translations and rotations. Bottom row: Spatial distribution of contact-point sliding and rolling. All values are normalized by their means, so that they can be represented with the same color scale. The moveable wall compresses from left to right, in the $+y$-direction. }
\end{figure*}

\emph{Data acquisition.---}Our experimental setup substantially upgrades the setup in Refs.~\cite{Peshkov:2019,Benson:2021}. Approximately 12,000 acrylic beads are placed in a square container with 15-cm-long sides. Each bead has a diameter of 5 mm and a mass of 0.0618 g.  The grains settle into the container to a height of 6 cm. We submerge the grains in a refractive-index-matched fluid (Triton X-100), in which a fluorescent dye (Nile Blue Perchlorate) is dissolved. A 1-kg acrylic weight is placed atop the grains, to apply a constant pressure. Sheet lasers on both sides of the container activate the fluorescent fluid in a two-dimensional plane. The nonfluorescent grains appear as dark regions. Moving the sheet lasers perpendicularly to the imaged plane (along the $z$-axis), we reconstruct a 3D image of the entire material. The grains have two perpendicular, cylindrical cavities. When a grain is immersed in the fluid, the cavities appear as a cross in the image (Fig.~\ref{fig:one}). The cross’s orientation shifts across sequential images, allowing us to calculate the grain’s rotation.

During the experiment, we cyclically compress the container along the $y$-axis with a compression amplitude of 0.15 cm (1\% of the container size). A compression cycle consists of 8 compression steps and 8 decompression steps.  At the end of each step, a 3D image is taken. The number of images taken across a cycle was chosen purely for convenience---to facilitate the tracking of the grains' positions and orientations. We compress the material at a speed of 0.001 cm/s---much less than the characteristic gravitational settling speed for the grains in the fluid~\cite{Dijksman:2012}. We compress the system in 400 consecutive cycles, to ensure that transient effects die out~\cite{Benson:2021}.

The Supplemental Materials detail the techniques used to infer each grain's position and orientation. To identify the positions, we use a modified Gaussian filter\cite{Harrington:2014} to identify candidate grains in each snapshot. To determine which candidates are real grains, we use a deep-learning model called a \emph{variational autoencoder}~\cite{Kingma:2013,Doersch:2016,Weishun:2021}. The model condenses the candidate grains' images into a two-dimensional latent space (Fig.~\ref{fig:two}). The neural network clearly separates the real and false candidate grains' latent representations. Once we infer a grain's position, we use a rotating Laplacian-of-a-Gaussian filter to identify the grain's orientation (Fig.~S1).

\textit{Rotational motions dominate the bulk dynamics.}---Figure~\ref{fig:three} shows how displacements---grain translations and rotations, as well as contact-point sliding and rolling---are distributed across the sample. Each displacement is calculated at full compression. For instance, a rotational displacement is how much a grain’s orientation changes between one compression cycle's beginning and centerpoint.  Grains crystallize at the bottom of the compression cell, as horizontal dots form strings along the $x$-axis in Fig.~\ref{fig:three}. Throughout the rest of the manuscript, we exclude the crystallized boundary from our calculations.

Well-defined shear zones are visible in the translations. In contrast, the rotations are distributed more uniformly, although larger rotations cluster near the compression wall. Greater sliding and rolling displacements cluster near the compression wall, too, but without being confined to the translational shear band. 

\begin{figure}[b]
\includegraphics[width=0.8\columnwidth]{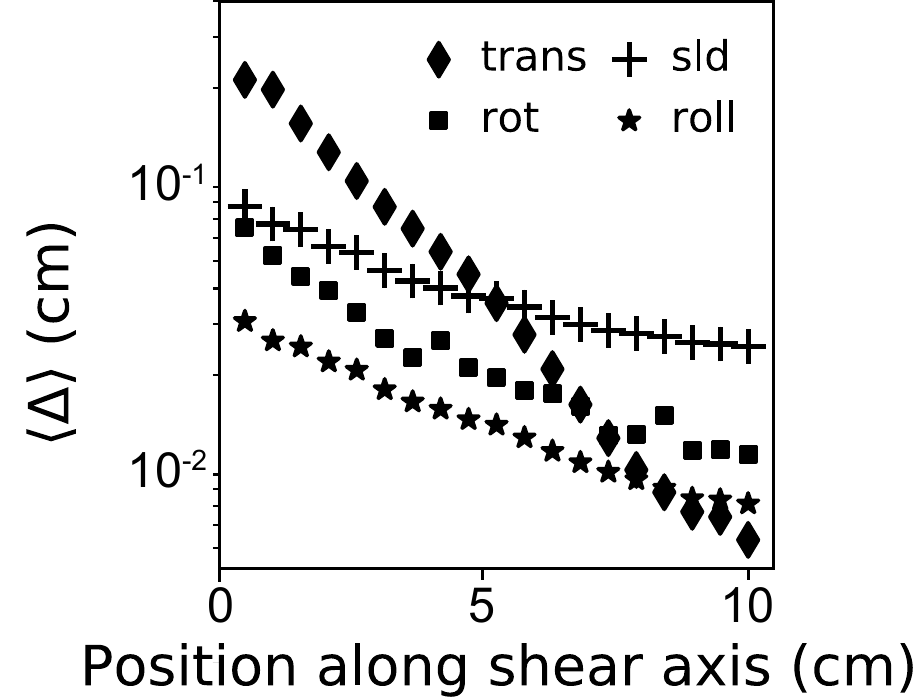}
\caption{\label{fig:four} Translations, rotations, sliding, and rolling as functions of position along the shear axis. Data are taken at full compression (from the middle of a cycle). The shear axis lies perpendicularly to the shear zone’s striations (see Fig.~\ref{fig:three}). The sliding and rolling displacements result from summing the displacements across the first half of the time steps.}
\end{figure}

The observations above invite us to quantify the spatial distribution of each type of motion. Continuum models suggest that rotational motion, modeled as a field, is proportional to the \emph{vorticity}---the spatial gradient of the translation field between shear layers~\cite{Muhlhuas:1987}. This result suggests that the rotations should be the spatial gradient of the translations. Since rotations are expected to couple to translations in the shear zone~\cite{Muhlhuas:1987}, we focus on this region. We quantify motions along a shear axis perpendicular to the shear zone (a 45-degree axis from the top left of Fig.~\ref{fig:three} to the bottom right), shown in Fig.~\ref{fig:four}. 

Along the shear axis,  translations diminish
more rapidly than rotations, sliding, and rolling. {\color{Purple} This result is expected, as the compression wall will perturb beads closer to the wall more than farther-away beads. We remove this bias in Figure S3. Here, we assume linear strain and subtract that component from the translations. Once adjusted, we find that the rotational displacements behave similarly to the translations. The difference between the two increases further into the bulk, where there appears to be more rotations. These excess rotations could be caused by the large amount of sliding in the bulk. } Two causes may explain the rotations' and contact-point motion's dominance. First, the sliding grains may be \emph{rattlers}, which carry no substantial loads and so may move relatively freely. Second, more motions generate rotations than generate translations: A translating grain can generate central contact forces, causing other grains to translate, and tangential forces, causing rotations. In contrast, frictional contacts' rotations can generate only tangential forces. {\color{Purple}However, due to the cavities, the beads in the experimental study are slightly non-spherical, allowing rotations to exert normal forces for some bead orientations. }

\begin{figure}[t]
\includegraphics[width=0.99\columnwidth]{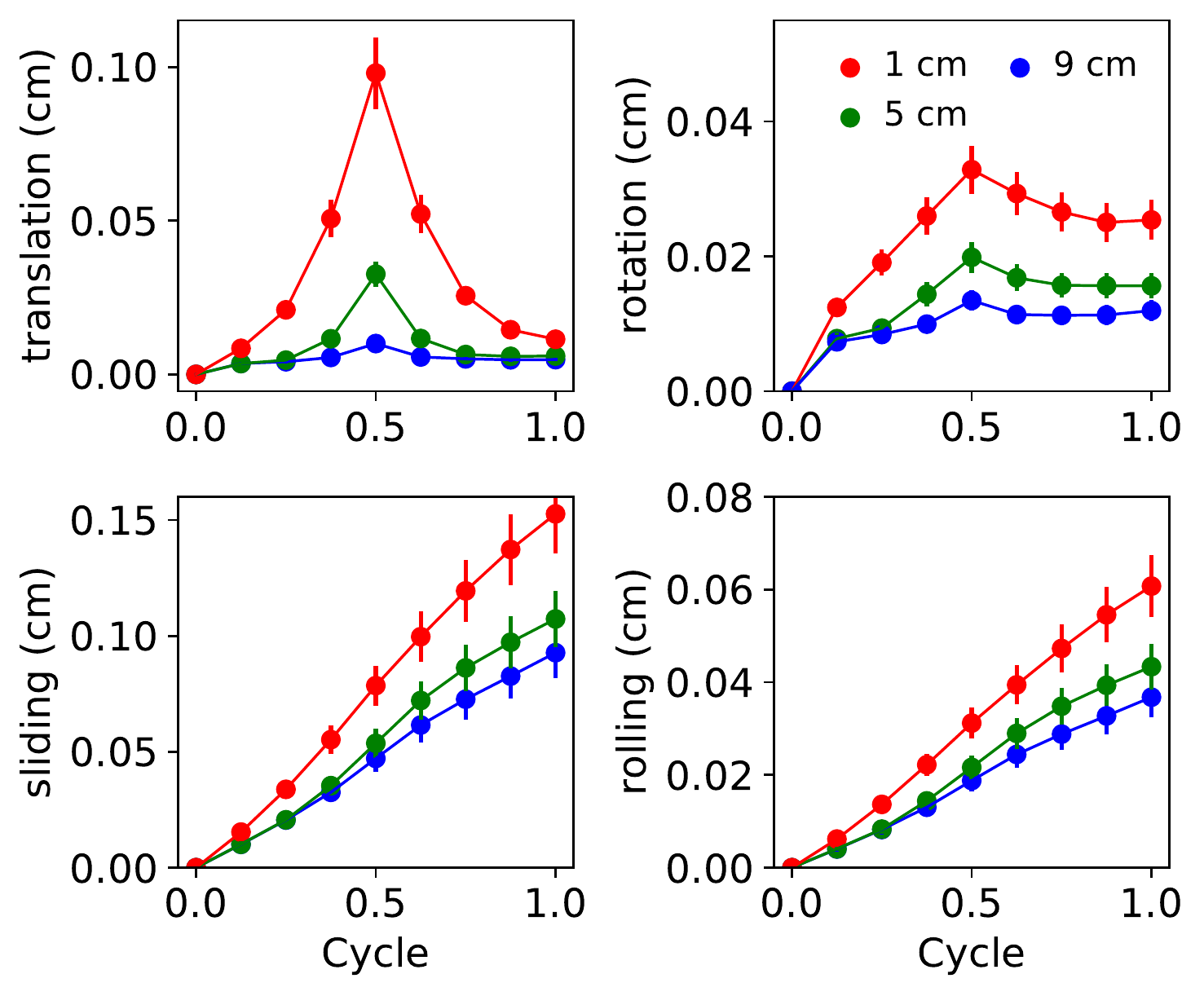}
\caption{\label{fig:five} Grains' mean translations and rotations (top row), as well as contact points' slipping and rolling (bottom row), across a cycle.  Along the horizontal axis are the steps within the cycle. Compression occurs during 0--0.5, and decompression occurs during 0.5--1.0. Color represents distance from the compression wall. Sliding and rolling are summed across a cycle.}
\end{figure}

\textit{ Reversibility of displacements}.---We have established that rotations and  sliding displacements penetrate more deeply into the bulk than translations do. We now quantify the different motions' reversibility across a compression cycle. Figure~\ref{fig:five} illustrates the four types of motion within a cycle, in three regions along the shear axis: (i) where translations occur more than rotations (red), (ii) where both motions occur to similar extents (green), and (iii) where rotations occur more (blue). Each successive region lies farther from the compression wall. We compute the dynamics at 9 instants within a cycle, then average each instant over cycles. In Fig.~\ref{fig:five}, the horizontal axis represents the fraction of a cycle that has passed. Compression occurs between points 0.0 and 0.5, and dilation occurs between 0.5 and 1.0. We calculate translations and rotations with respect to the beginning of the cycle. We sum sliding and rolling displacements between consecutive cycle steps.

A few key features merit discussion. First, the mean 3D rotation is more irreversible than the mean translation, consistent with findings limited to 2D rotations~\cite{Peshkov:2019,Benson:2021}: The mean translation (top left) grows during compression, then decreases during decompression. In contrast, the mean rotations (top right) are asymmetric with respect to compression and dilation. The mean rotation grows during compression, then changes little during decompression. Hence the mean 3D rotation is not reversible, although the mean translation is. This conclusion holds close to and far from the compression wall, as evidenced by points of all colors.

The bottom row of Fig.~\ref{fig:five} shows the mean contact-point dynamics. Throughout the cycle, the average sliding and rolling displacements grow. However, consistent with Fig.~\ref{fig:four}, the mean sliding and rolling diminish as the distance from the compression wall increases. As sliding dissipates heat, our observations of granular motions merit comparison with energy dissipation.

\textit{Energy dissipation.}---Under quasistatic conditions, dense granular materials dissipate energy primarily through frictional contacts that slide~\cite{Zhai:2019,Dietrich:1998}. We can measure sliding using the experimental and analytical tools introduced in this paper.  Calculating energy dissipation, though, requires also knowledge of contact forces. Thus, we evaluate dissipation using numerical simulations detailed in the Supplemental Materials~\cite{Schwartz:2012}. Figure~S4 shows the analog, produced via simulation, of Fig.~\ref{fig:five}. The simulations and experimental data agree excellently, allowing us to infer from the simulations how the granular material dissipates energy.

\begin{figure}[ht]
\includegraphics[width=0.8\columnwidth]{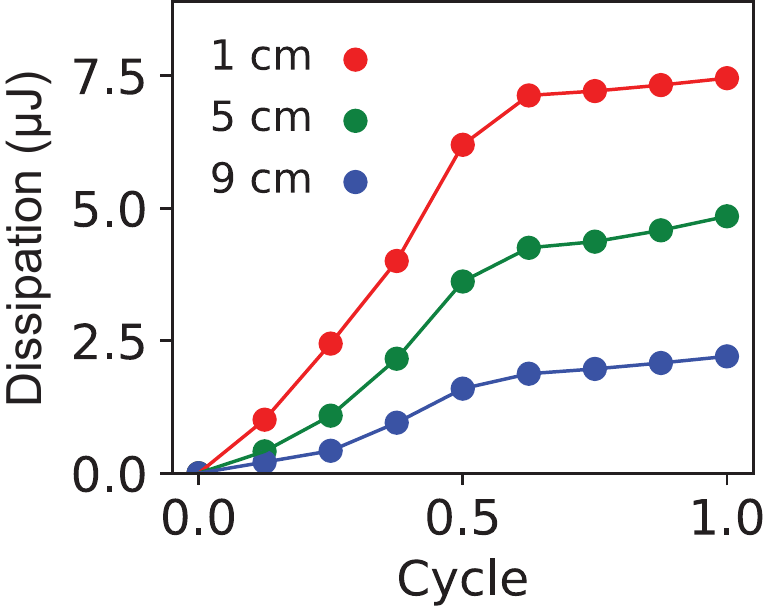}
\caption{\label{fig:six} Cumulative energy dissipated via friction throughout a given fraction of the cycle. Different colors represent different distances from the compression wall. The dissipation happens primarily during compression (0.0--0.5).
}
\end{figure}

Figure~\ref{fig:six} shows the energy dissipation accumulated across the cycle, across the material's three regions. The material dissipates energy primarily during compression, as the curve flattens during decompression. Hence the energy dissipation is asymmetric across the cycle, similarly to the mean rotation
~\footnote{
One might expect the material to dissipate substantial energy during decompression,
because (i) the grains slide substantially during decompression (Fig.~\ref{fig:five}) and (ii) sliding tends to dissipate energy. However, the dissipation depends not only on the sliding distance, but also on the force at the contact point. This force is smaller, on average, for larger sliding displacements (Fig.~S5).
}.
Furthermore, the dissipation is significant deep in the sample, where the grains are rotating more than translating (Fig.~\ref{fig:four}). These observations suggest that 3D rotations, rather than translations, dominate frictional energy loss, even deep in the bulk.

 \emph{Conclusions.---}We have experimentally measured 3D rotations of spheres in dense granular packings. In addition to measuring each sphere's rotations, we quantified the rolling and sliding at contact points between grains. We have discovered that rotations and contact-point motions penetrate further into the granular material than translations do. Throughout the system, rotations are irreversible. As a result, sliding displacements accumulate throughout the cycle. Our experiment points to the accuracy of numerical simulations from which we estimated energy dissipation. We find that dissipation occurs deep in the sample, where grains rotate more than they translate. This result suggests that energy is transferred and dissipated far from the perturbation's source through rotations, not through compressive forces from collisions caused by translations alone. Our work establishes a rich foundation for experimental measurements of 3D rotations and demonstrates that rotations are crucial for understanding granular materials' unique energy-dissipation characteristics.

\emph{Acknowledgements.---}This work was supported by University of Maryland supercomputing resources~\footnote{http://hpcc.umd.edu}. Z.A.B. was supported by the National Science Foundation graduate research fellowship program. N.Y.H. is grateful for an NSF grant for the Institute for Theoretical Atomic, Molecular, and Optical Physics at Harvard University and the Smithsonian Astrophysical Observatory, as well as for administrative support from the MIT CTP. WL and AP were supported by NSF No. DMR-1507964. AP work was supported by NSF Grant No. DMR-1809318.

\bibliographystyle{aipnum4-2}
\bibliography{references_database}

\end{document}